\documentclass{PoS}
\newcommand{\hi}{{\sc H\,i} }

\title{Gravitationally Lensed \hi with MeerKAT}

\ShortTitle{Gravitationally Lensed HI with MeerKAT}

\author{\speaker{Roger Deane}%
        \\
       Rhodes University\\
       E-mail: \email{r.deane@ru.ac.za}}

\author{Danail Obreschkow\\
        ICRAR\\
        E-mail: \email{danail.obreschkow@gmail.com}}
\author{Ian Heywood\\
        University of Oxford, Rhodes University\\
        E-mail: \email{ian.heywood@gmail.com}}

\abstract{The SKA era is set to revolutionize our understanding of neutral hydrogen ({\sc H\,i}) in individual galaxies out to redshifts of $z\sim0.8$; and in the $z > 6$ intergalactic medium through the detection and imaging of cosmic reionization. Direct \hi number density constraints will, nonetheless, remain relatively weak out to cosmic noon ($z\sim2$) - the epoch of peak star formation and black hole accretion - and beyond. However, as was demonstrated from the 1990s with molecular line observations, this can be overcome by utilising the natural amplification afforded by strong gravitational lensing, which results in an effective increase in integration time by the square of the total magnification ($\mu^2$) for an unresolved source. Here we outline how a dedicated lensed \hi survey will leverage MeerKAT's high sensitivity, frequency coverage, large instantaneous bandwidth, and high dynamic range imaging to enable a lasting legacy of high-redshift \hi emission detections well into the SKA era. This survey will not only provide high-impact, rapid-turnaround MeerKAT science commissioning results, but also unveil Milky Way-like systems towards cosmic noon which is not possible with any other SKA precursors/pathfinders. 
 An ambitious lensed \hi survey will therefore make a significant impact from MeerKAT commissioning all the way through to the full SKA era, and provide a more complete picture of the \hi history of the Universe.}

\FullConference{MeerKAT Science: On the Pathway to the SKA,\\
	25-27 May, 2016,\\
		Stellenbosch, South Africa}

\newcommand{\skipthis}[1]{}
\newcommand{\mnras}{MNRAS}
\newcommand{\nat}{Nature}

\newcommand{\apj}{ApJ}
\newcommand{\apjl}{ApJL}
\newcommand{\aap}{A\&A}

\newcommand{\araa}{ARA\&A}
\newcommand{\pasj}{PASJ}

\newcommand{\nar}{New Ast. Reviews}
\newcommand{\physrep}{Physical Reports}

\begin{document}

\section{Introduction}
The evolution of neutral hydrogen is fundamental to our understanding of galaxy evolution. Our current cosmological picture is that following recombination at $z\sim1100$, the Universe was predominately neutral until reionisation occurred between $z\sim6-10$, driven by accretion onto supermassive black holes and star formation in the first galactic haloes. Followed by what the current data argue to be a relatively rapid reionoization epoch between $z \sim 6-7$ (e.g. \cite{Robertson2013}), most neutral hydrogen is now found in galaxies where it is shielded and/or replenished (e.g. \cite{Barkana2001, Fan2006}). However, the evolution of neutral hydrogen in $z<6$ galaxies is poorly constrained, as direct detections are only routinely possible out to $z \lesssim 0.2$. Limits from Ly-$\alpha$ absorbers suggest little to no evolution, albeit with large systematic and statistical uncertainties (e.g. \cite{Prochaska2009}), while statistical techniques (i.e. stacking and intensity mapping; e.g. \cite{Lah2007, Rhee2016, Chang2010}) are yet to yield robust, high precision results beyond low redshifts (e.g. \cite{Delhaize2013}). 

%should mention CHILES z~0.376 detection somewhere

The above, data-limited picture is set to change drastically over the next 5-10 years, which will fundamentally transform our understanding of the cosmic evolution of neutral hydrogen. This will be driven by a wide range of new radio telescopes, including MeerKAT, the Australian Square Kilometre Array Pathfinder (ASKAP), SKA1-MID, as well the upgraded Karl G. Jansky Very Large Array (VLA), Westerbork Synthesis Radio Telescope (WSRT) with the APERture Tile In Focus project (APERTIF), and the Giant Metrewave Radio Telescope (GMRT). These telescopes will perform multi-tiered \hi~surveys (an indicative summary can be found in \cite{Blyth2015}), which will increase \hi sample sizes by several orders of magnitude. While the deepest of these surveys will detect individual galaxies out to redshifts of $z\sim1$ for the most massive \hi systems ($M_{\rm HI} \gtrsim 10^{11}$~M$_{\odot}$), all will have \hi limiting masses well above a Milky Way-like system at $z\gtrsim0.8$. Higher redshifts will be accessible through \hi~absorption detections which will provide important insights (e.g. \cite{Gupta2006, Allison2015}), however, these still suffer from significant sources of uncertainty (e.g. spin temperature, gas morphology) that limit extrapolation to the cosmic evolution of \hi~in galaxies.

The greatly enhanced sensitivity, survey speed and instantaneous \hi redshift coverage of these next-generation instruments will result in a dramatically increased {\sc H\,i}-detection probability of both known and unknown lensed sources. Here we assume the canonical definition of strong lensing as a total magnification $\mu\geq2$. This corresponds to an effective increase in integration time by a factor of $\mu^2\geq4$, assuming the lensed sources are unresolved in the image plane. Since routine \hi detection have so far been limited to $z \lesssim 0.2$ ($D_{\rm L} \sim 980$~Mpc), the chance alignment of two unrelated galaxies within this volume is negligibly small. However, since the chance alignment probability goes as $r_{\rm max}^{6}$ for a Euclidean non-evolving Universe, where $r_{\rm max}$ is the distance that corresponds to the maximum redshift, the lensing probability is set to dramatically increase as \hi sensitivity limits increase to $z \sim 1$ ($D_{\rm L} \sim 6\,800$~Mpc). The opening of the \hi Universe by these future radio facilities therefore opens an additional window through lensing and enables the detection of \hi at even greater cosmological distances than otherwise expected. %with future instruments.

The two highest priority MeerKAT science cases are a deep \hi survey and pulsar timing (Baker et al., Bailes et al., these proceedings). However, what had not previously been considered in the MeerKAT science case, is the enhanced ability to detect \hi emission in high-redshift galaxies by using the natural amplification of gravitational lensing. In this contribution, we consider the scientific opportunities this would provide and the niche that MeerKAT will occupy in this domain until the SKA-era. We discuss possible observing strategies to maximise of the scientific return of lensed \hi detections and outline how this will enhance \hi science with MeerKAT. We also outline how gravitational lensing will enable high-impact, rapid-turnaround early science with MeerKAT and provide a preview of the revolutionary impact the LADUMA and MIGHTEE-HI surveys will have, once completed.

\section{HI surveys in the MeerKAT and SKA era}

\begin{figure}[b]
\centering
\includegraphics[width=0.6\textwidth]{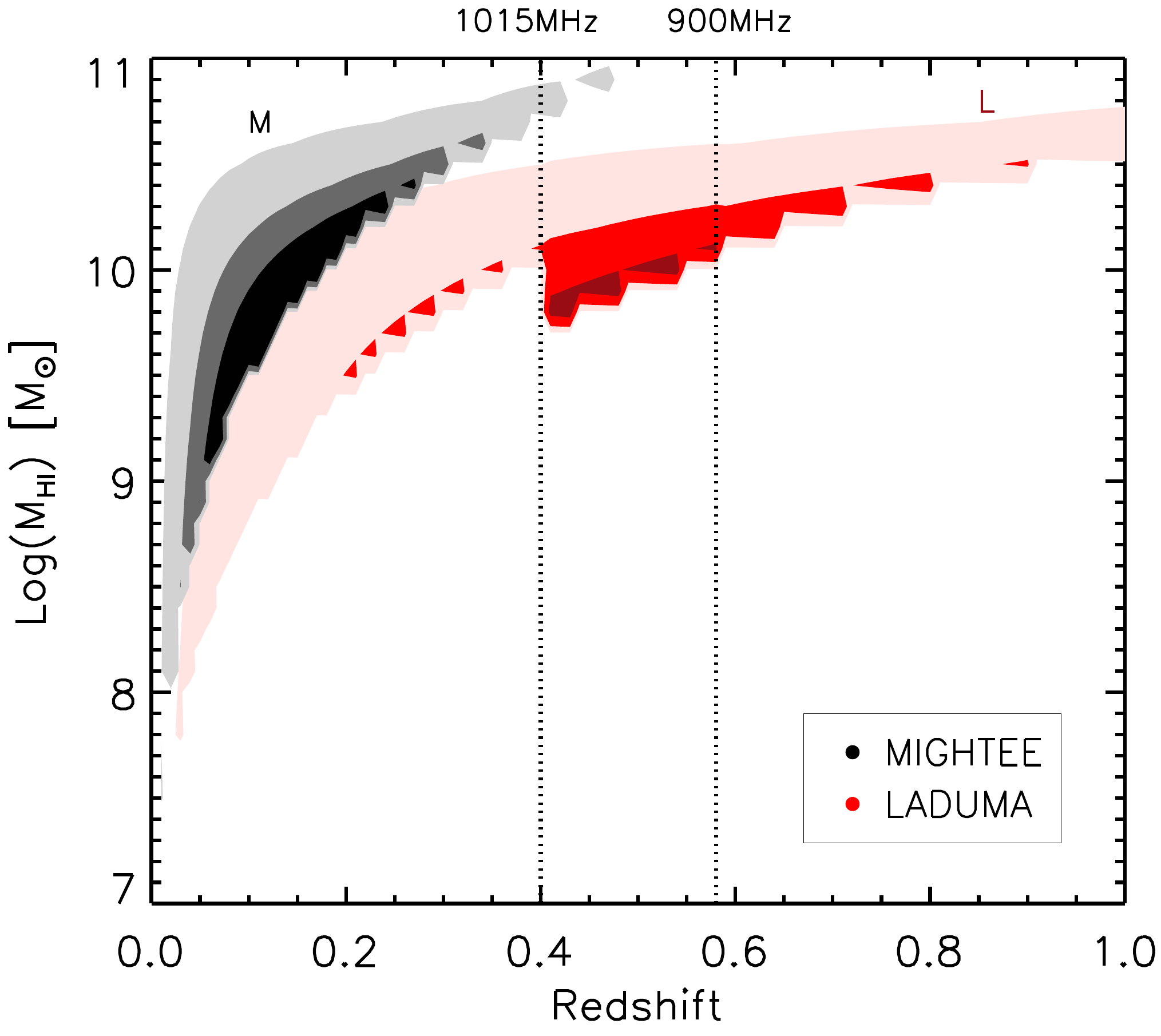}
\caption{Directly reproduced from Figure 2 of \cite{Maddox2016}.The coverage in the redshift-M$_{\rm HI}$ plane for LADUMA and MIGHTEE, demonstrating the lack of direct detections beyond $z\gtrsim 1$, which lensing could enable. The parameter space is divided into cells of width 0.01 in redshift and 0.1 in $\log{(M_{\rm HI})}$. The contour levels for both surveys are the same, surrounding regions of parameter space containing at least one (faint shading), 10 (medium shading), and 25 (dark shading) galaxies per cell. Due to the overlapping frequency coverage of the two MeerKAT receivers, the redshift range $0.40 < z < 0.58$ is observed for 5000 hours for LADUMA, and is correspondingly deeper.}
\label{fig:maddox}
\end{figure}

A primary objective of the SKA from its earliest developments was to detect Milky Way-like systems in \hi emission at high redshift ($z\sim2$; \cite{Taylor1999}). %As the science case evolved, the key science project emerged to detect Baryonic Acoustic  Oscillations (BAO) to high precision in order measure the dark energy equation of state (Abdalla \& Rawlings 2005), which would require the full SKA to detect $\sim10^9$ \hi galaxies at redshifts up to $z\sim 1.5$. 
This objective has evolved into a rich science case for {\sc H\,i}, covering many key aspects of galaxy evolution physics and cosmology from the detection of the epoch of reionization (\cite{Santos2015}), to Baryonic Acoustic  Oscillations (BAO) studies (\cite{Rawlings2004}), and potential direct expansion measurements (\cite{Kloeckner2015}). Current specifications of SKA1-MID Band 1 would only enable detections above $z>0.8$ for $M_{\rm HI} \gtrsim 10^{10}$~M$_\odot$ \hi galaxies in the {\sl deep}, 1~deg$^2$ survey outlined by \cite{Staveley-Smith2015} and SKA Science Book chapters referred to in their overview. They describe three fiducial SKA1-MID \hi surveys that cover extragalactic fields: the {\sl medium wide}, {\sl medium deep} and {\sl deep} surveys which cover areas of $\Omega$ = 400, 20, 1 deg$^2$ respectively, the first two of which are limited to Band 2 and its corresponding \hi redshift limit of $z<0.5$. In addition to this, there may be a 10,000 hour SKA1-MID survey, with a commensal \hi cosmology and galaxy evolution design, however, the survey design is yet to be determined (see \cite{Staveley-Smith2015}). 

%[xx might need to add line on complementary 10,000 hour surveys with SKA1-MID (table 2 in SS\&O16)].  

From an SKA precursor/pathfinder perspective, the only survey capable of detecting \hi emission beyond $z\gtrsim 0.6$ will be the dedicated, several thousand hour MeerKAT LADUMA \hi survey - a single pointing (0.6~deg$^2$ at $z = 0$) covering the Extended Chandra Deep Field South (ECDFS) using both L- and UHF bands (Baker et al., these proceedings). This will have strong complementarity with the spectral line component of the MIGHTEE survey, which will cover an area of $\Omega \sim 20$~deg$^2$ and hence provide better constraints on the \hi mass function at the high end out to a redshift of $z < 0.58$ \cite{Maddox2016}. 

While none of these \hi surveys will be able to detect Milky Way-like galaxies at $z\sim1$, a major opportunity to do so lies in the use of strong gravitational lensing. The significant increase in sensitivity and redshift space accessible by next-generation facilities, exemplified by SKA1-MID and MeerKAT, will ensure the routine detection of lensed \hi systems. To take advantage of this opportunity, we have performed detailed lensed \hi simulations, that rely on sophisticated N-body simulations over a large cosmological volume, employing full general relativistic ray tracing, which we briefly review in the following section.

%mostly to probe galaxy evolution in a wide range of environments. 

%add a table? based on Lister-Tom, Danail, Sarah. 

\section{Semi-analytic lensed HI simulation}

%To address these questions as accurately as possible, we carried out an N-body simulation with realistic semi-analytic prescriptions for the spatially-extended \hi disks (by size, density profile and rotation), combined with general relativistic ray tracing (for individual frequency channels), to predict the magnification statistics and resultant number counts of strongly-lensed \hi emission. 

Given the opportunity \hi lensing presents, \cite{Deane2015} designed a simulation to address several questions, including:
what are the {\it survey-independent} \hi lensing statistics as a function of redshift? How will SKA (precursor/pathfinder) surveys perform in making detections? In the absence of a pre-defined survey, which telescope is best suited to detecting lensed systems? What is the actual scientific utility of detecting lensed \hi systems? To address these questions, we used a semi-analytic model producing realistic \hi disks in terms of size, density profiles and rotation structure. This model was run on the Millennium simulation of cosmic structure, enabling us to account for realistic galaxy-galaxy clustering when computing the alignment of foreground and background galaxies. For sufficiently aligned pairs, realistic {\sc H\,i}-lensing factors were computing using general relativistic ray tracing, applied individually to each frequency channel (see Fig.~\ref{fig:chanmaps}). Our simulation uses a 150~deg$^2$ field-of-view mock observing cone out to a redshift limit of $z_{\rm source}<4$, enabling the prediction of detection rates of virtually all future \hi surveys (excluding SKA-LOW), for intrinsic velocity-integrated \hi flux densities above 0.1 mJy\,km\,s$^{-1}$. 
 
These simulations resulted in a wide range of conclusions, two of which are particularly relevant to these proceedings:
 \begin{itemize}
 
 %\item Efficient lens selection will only be possible in wide area SKA1-MID surveys, particularly beyond $z\gtrsim 2$ due to the steepness of the \hi luminosity function at the high-mass end.
 \item[-] None of the SKA pathfinder/precursor wide-field surveys will detect a large number of \hi lenses, apart from MeerKAT's LADUMA survey ($\sim$20 at $5\sigma$), however, the vast majority of those will have low magnification ($\mu \sim 2$) with limited scientific utility. 
 \item[-] MeerKAT is the best-suited instrument to make lensed \hi detections pre-SKA, provided it performs a targeted survey of known gravitational lenses, as well as blind, single pointing searches towards well-modelled, high mass clusters with minimal associated diffuse radio emission.
 
 \end{itemize}

In the remainder of this contribution, we focus on MeerKAT-specific technical advantages and scientific opportunities with regards to detecting lensed \hi systems.

%{\bf Figure 2: overall lensing stats as func of redshift? with telescope redshift cutoffs indicated. }

\section{MeerKAT's design advantages for lensed \hi detection}\label{sec:design}

MeerKAT's design was optimised toward its two highest priority science themes: neutral hydrogen beyond the local Universe and pulsar searching and timing. A few of the resultant telescope attributes are a `pinched' core array configuration (50\% of the collecting area within 1 km)\footnote{A pinched core refers to an array configuration that has a significant increase in filling factor within the core, thereby achieving high brightness temperature sensitivity and pulsar search efficiency.}; excellent receiver sensitivity in the L-  and UHF bands; wide field-of-view ($\sim$2 and 0.6~deg$^2$ for UHF and L-band respectively); wide bandwidth (435 and 770 MHz UHF and L-band respectively); and an offset-Gregorian feed for completely unblocked aperture to maximise the imaging dynamic range. The array configuration will result in high fidelity imaging and the ability to accurately model bright sources, while still not achieving too high an angular resolution as to begin to resolve out strong lenses with their increased solid angles. However, even in the case that this does occur - the pinched core array configuration will mean that just a small fraction of the total sensitivity is lost to the highest magnification cases.

 If one combines the above with the extremely radio quiet site, as well as the imaging quality and measured phase stability from MeerKAT Array Release 1 (AR1), this suggests MeerKAT is supremely placed to detect high redshift, gravitationally lensed \hi emission. The combination of the wide FoV and large instantaneous bandwidth mean a large cosmological volume (co-moving volume of $V \sim 15$~Gpc$^3$ in UHF band) is observed for every pointing. This maximises the probability of serendipitous detections in the case of blind cluster surveys and enables rich commensal science opportunities.

\section{Science opportunities enabled by gravitational lensing}

Although gravitational lensing has a long history of providing a rich set of results at other wavelengths, it is important to ask what the specific scientific utility will be in the lensed \hi case. Here we describe a just a few of the wide range of science applications, all of which have significant value and will further the MeerKAT \hi science priority.

\subsection{Highest redshifts}\label{sec:highz}

Amongst the most spectacular results from MeerKAT is likely to be the direct detection of lensed, gas-rich \hi galaxies at the highest redshifts achievable in the pre-SKA era ($z\sim1.45$). Simulations predict that LADUMA will only be able to detect a $M_{\rm HI}\sim10^{10}$~M$_{\odot}$ galaxy out to $z\sim1$ in its $\sim$2500~hours single UHF band pointing (see Baker et al., these proceedings; \cite{Maddox2016}). However, for a magnification $\mu$ = 10, 25 system, this would correspond to an effective factor of $10^2$ = 100, $25^2$ = 625 increase in observing time. So for the latter case, a 24 hour MeerKAT observation would be the equivalent of $\sim8 \times$ LADUMA for that particular galaxy. Clearly, this will provide the deepest cosmic view of \hi emission in galaxies. Even this high magnification system would still be within 2-3 PSFs, assuming a $r \sim$30 kpc disk at $z\sim1$, which is only magnified in one direction (i.e. worst case scenario, which is highly unlikely given the expected, intrinsic \hi solid angles). 

The key scientific applications here are to (a) reconcile the dramatic increase in cosmic star formation rate density (\cite{Madau2014}, and references therein) with the relatively constant \hi density evolution (e.g. \cite{Prochaska2009}); and (b) obtain a high-redshift \hi perspective on the dichotomy between so-called main sequence star forming galaxies which are driven by secular evolution (e.g. \cite{Daddi2007}) and the starbursts that are driven by major mergers (e.g. \cite{DiMatteo2007}). Given the three order of magnitude increase in the cosmic star formation rate density and the significantly higher galaxy merger rate at $z\sim2$, direct \hi emission measurements at the highest redshifts are of key importance and likely to only be delivered through strong gravitational lensing. 

These detections at the highest redshifts will be prime targets for followup with millimetre telescopes into order to compare the global kinematics and measure the cosmic evolution of the molecular-to-atomic hydrogen ratio in galaxies. The latter is of particular importance in the understanding of the star formation history of the universe. Theoretical models predict a significant decrease in the {\sc H}$_2$/\hi ratio as a function of cosmic time, which is thought to be driven by the growth of galactic disks and the systematic decrease in the disk mid-plane cold-gas pressure (e.g. \cite{Obreschkow2009}). Since CO transitions are readily detectable at these redshifts, it is direct \hi detections that that will limit high redshift constraints of the molecular-to-atomic hydrogen ratio out to $z\gtrsim1$. Futhermore, molecular line follow up is likely to be an efficient way to confirm \hi lens candidate redshifts, particularly for the blind cluster surveys outlined in Sec.~\ref{sec:cluster}, which may have OH megamaser contaminants (e.g. \cite{Briggs1998}).

\subsection{Low \hi mass systems at cosmological distances}

As shown by our simulations in \cite{Deane2015}, two factors drive the dramatic increase in probability that lower mass \hi galaxies ($\lesssim10^9$~M$_\odot$) are strongly lensed: (a) their increased space density, and (b) their smaller solid angles with respect to higher mass systems, which for \hi disks typically results in higher magnification factors (all other factors being equal). What this implies is that MeerKAT's specifications make it extremely well suited to both targeted and serendipitous detections of lower mass \hi systems across a wide range of redshift space ($0.2 \lesssim z \lesssim 0.6$). Furthermore, it may even  discover extremely high magnification systems ($\mu > 50$) that have intrinsically small solid angles and lie directly on a gravitational lens caustic or cusp. These possibilities open up the opportunity for detailed study of lower mass \hi galaxies simply not possible with any other \hi surveys that are currently planned. 

While it is more difficult to discern high magnification, low mass galaxies from high mass, non-lensed galaxies at similar redshifts, these are in principle easily to distinguish though cross-matching with foreground, low impact factor galaxies, particularly bright ellipticals which are easily identifiable from optical/NIR imaging.

%\subsection{Probing foreground galaxies and dark matter haloes}
\subsection{Efficient lens selection and a unique probe of foreground dark matter haloes}\label{sec:lensselect}

\begin{figure}
\centering
\includegraphics[width=0.7\textwidth]{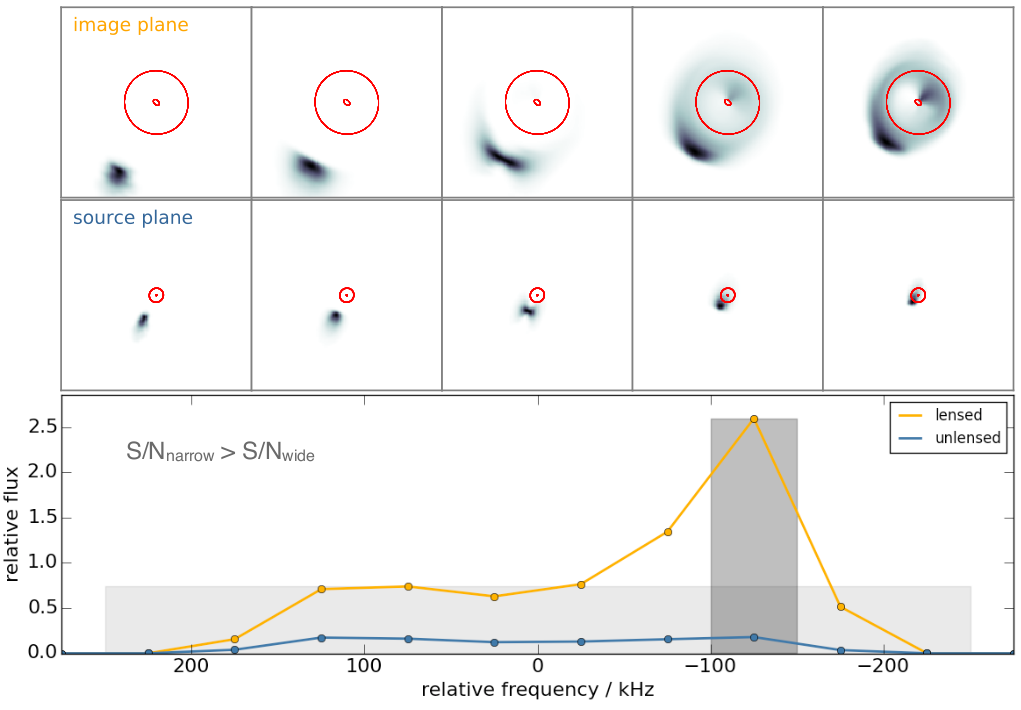}
\caption{Directly reproduced from \cite{Deane2015}. {\bf Top: }Image and source plane channel maps of a Milky Way mass \hi disk (frequency resolution = 200 kHz), demonstrating how the approaching \hi emission overlaps with the inner lens caustic and results in a full \hi Einstein ring in some channels.  Each frame is 6.4$\times$6.4~arcsec$^2$. The caustics/critical curves are plotted in red. {\bf Bottom: }Lensed and unlensed \hi profiles of the same system at higher frequency resolution (50 kHz). Differential magnification is seen between channels to the extent that the probability of detection is higher for a narrow ($\sim$50 kHz) channel width, rather than the full \hi line width. }
\label{fig:chanmaps}
\end{figure}

Two properties of \hi lends itself to efficient gravitational lens selection, provided the appropriate survey is designed. Firstly, the redshifted \hi line is very isolated in the frequency range $580$~MHz~$< \nu < 1420$~MHz ($1.45 > z > 0$), although OH megamasers may prove to be a science-rich contaminant \cite{Briggs1998}. Secondly, the high end of the \hi mass function is very steep ($n \propto s^{-3}$). This means that deep \hi observations can apply a very similar lens selection technique to that performed at far-infrared/(sub)mm wavelengths in the \emph{Herschel}-ATLAS survey \cite{Negrello2010}, and with the South Pole Telescope, both of which had SMA/ALMA/HST followup with spectacular results \cite{Vieira2013}. The basic idea is that given the extreme rarity of the highest luminosity systems and the steep drop off in number counts, any object that deviates significantly from the statistical number density expectation is likely to be lensed. Following the removal of some obvious contaminants (blazars and Galactic objects), \cite{Negrello2010} were able to achieve an unprecedented $\sim$100 percent lens selection rate. \hi surveys could apply the same technique, however, with the huge advantage of spectral line capability (i.e. a 3D rather than 2D selection). This would make cross-identification easier and also open the opportunity to discover systems that have a few velocity channels that have extremely high magnification factors. In the latter case, the detection probability may be larger for a channel search width significantly smaller than the galaxy's full velocity width, as demonstrated in Fig.~\ref{fig:chanmaps}. In addition to this, MeerKAT's design and sensitivity may enable the simultaneous detection of the foreground lens in either \hi emission and/or absorption. 

What this all means is extremely high efficiency lens selection and redshift measurement, which equates to scientifically-rich samples to study foreground dark matter halo statistics. While large volume SKA \hi surveys will be required to take full advantage of this opportunity, MeerKAT will undoubtedly pave the way. 

If the foreground lens galaxy is sufficiently low redshift ($z\lesssim 0.15$), gas-rich, with dynamics dominated by rotation, it could provide the ability to derive a dark matter mass model at large radii. If combined with optical/NIR resolved lensed images which constrain the mass density at smaller galactic radii, this could provide a unique probe of the galaxy mass profile at large radii. It is therefore quite possible that \hi will continue to provide insights in galactic dark matter content out to higher redshifts, just as it is done in the local Universe for the past few decades. 

%Need citation and application. Could something be done with dark matter halo statistics at lower z? Point is: lensing is unbiased mass estimator: so if you select many lenses in the same way at different redshifts, and you assume something about the HI mass function evolution, then sure you can constraint DM halo mass function evolution? Or at least put constraints? 

%This chapter and MeerKAT is focused on high-z HI, but really this DM science could be a huge yield and so the path towards SKA. 

\subsection{Enhanced spatial resolution to higher redshift}

MeerKAT's array configuration has maximum baselines of 8~km. With its pinched core array configuration, the angular resolution ranges between 10-15~arcsec (at 1.4~GHz) for Briggs-weighted images with robust parameters from 0.5-2. This means that a Milky Way type \hi disk will be unresolved beyond redshifts of $z\gtrsim0.15$. However, the largest Einstein radii of massive clusters ($\theta_{\rm Ein} \gtrsim 20$~arcsec, e.g. \cite{Oguri2009, Waizmann2014}), would result in a comparable angular extent magnification of an \hi galaxy and hence be spatially resolved. While some sensitivity would be lost to resolved sources, if the lensed source is sufficiently H\,{\sc i}-rich then it would enable the only method to investigate the resolved \hi kinematics of MeerKAT-detected sources beyond $z\gtrsim0.15$. Detailed modelling of spatially-resolved lensed molecular gas has revealed great insights into disk formation, merger activity, and dark matter content (e.g. \cite{Hodge2012, Deane2013}). Doing the same for the more extended \hi gas is an important systematic study to pursue with MeerKAT, which the highest mass clusters could enable, even if for just a handful of targets.

% comment to Ian on Waizmann: clusters get more massive at lower-z, so yes. I don't really want to get into too much detail.

\subsection{Probing the global kinematics of galaxies at high redshift}

Lensed {\sc Hi}-line detections will constrain the maximum velocity width ($V_{\rm max}$), which, in combination with a sufficiently deep adaptive optics or space-based NIR/optical imaging, allows a measurement of the baryonic angular momentum. The larger extent of \hi in galaxies means that these lensed \hi detections are likely to become the best baryonic angular momentum measurements at $z\sim1$, adding an important evolutionary perspective to the science case presented in \emph{The SKA as a doorway to angular momentum} \cite{Obreschkow2015}.

\section{MeerKAT lensed HI survey strategy}

There are a range of strategies that a dedicated lensed \hi survey with MeerKAT could take. Here we outline some considerations to make in formulating an observational programme that includes a proof-of-concept stage, followed by dedicated targeted observations of known lenses (selected at other wavelengths); and deep, high-mass cluster observations to enable serendipitous detections.

\subsection{Proof-of-concept / science commissioning observations}

The MeerKAT Large Survey Programmes (LSPs) have been allocated $\sim$70~percent of the observing time on MeerKAT during the $\sim$5~years it will operate before becoming part of SKA1-MID. Many of the approved LSPs require the full array (Array Release 6) to achieve their scientific objectives (laid out in these proceedings). While the LADUMA and MIGHTEE LSPs will require pre-AR6 time to test data post-processing pipelines and familiarise survey team members with the data attributes, it's not clear that there will be any high impact, quick turnaround results from these in the pre-AR6 stage. 

In contrast, a lensed \hi detection at a redshift inaccessible by current radio interferometers such as the VLA ($z \gtrsim 0.4$) would provide a high impact science result with just a few tens of hours of observing time (i.e. $\lesssim$50 $\mu$Jy\,beam per $\sim$200 kHz channel sensitivity at $\sim$1 GHz). This would simultaneously provide an excellent science commissioning target in both the bottom end of the L-band and top end of the UHF band. A handful of moderate redshift ($0.2 \lesssim z \lesssim 0.58$) detections would play an important role in demonstrating the ability to detect \hi at cosmological distances in just a few tens of hours, while avoiding the calibration risks and data-processing overhead of a few hundred hour commissioning observation. 

Regarding data volumes, there are clear advantages here. With observations of 24-48 hours, the data volume in the Measurement Set format is of the order of 72-144 TB (assuming 32K spectral line mode, 64 antennas, full polarisation), which means the full visibility dataset can be processed rather than being forced to combine subsets of the data in the image domain (a calibration risk for LADUMA, for example). During science commissioning, this data rate can be reduced by well over an order of magnitude by only processing a selected channel range set by the known redshift of the targeted lensed source. This demonstrates that lensed \hi targets can indeed deliver rapid-turnaround, high impact science-commissioning results with relatively low calibration and post-processing risk and expense. 

Lensed sources have been shown to be excellent science commissioning targets (or byproducts) in other observatories, most recently by \emph{Herschel} and ALMA. For example, discoveries in the \emph{Herschel}-ATLAS Science Demonstration Phase included the $z \sim 3$ Einstein-ring source SDP.81 using the technique described in Sec.~\ref{sec:lensselect} (Negrello et al. 2010), which was then followed up to produce the spectacular images seen in ALMA Science Verification 2014 Long Baseline Campaign. These commissioning observations resulted in a number of publications of this single lensed system (e.g. \cite{ALMA2015, Dye2015, Rybak2015, Swinbank2015, Tamura2015, Hezaveh2016}). MeerKAT has a similar opportunity to use lensing to detect \hi at unprecedented redshifts while still in its science commissioning phase, the followup study of which will undoubtedly yield rich scientific rewards.

\subsection{Targeted surveys}\label{sec:targeted}

Following the proof-of-concept detections, the next step would be to perform a targeted survey of a list of known lenses that are detected at other wavelengths. In the redshift range accessible to MeerKAT, the vast majority of known lenses (at present) were discovered in the SLACS survey (\cite{Bolton2006}), which identified composite (i.e. emission + absorption lines), but redshift-offset SDSS-spectra and performed high angular resolution HST followup to identify the largest sample of lensed sources to date (this includes 46 sources below MeerKAT's maximum \hi redshift limit of 1.45). This preferentially selected extended star forming galaxies lensed by foreground elliptical galaxies. There are a number of other lenses in this redshift range, selected in a range of methods (CLASS, SQLS, COSMOS, GAMA; see \cite{Treu2010} and reference therein), however, they make up the minority, at present. 

The most promising lensed \hi targets within these samples can be identified using the (large scatter) $M_{\rm HI}$-$M_{\star}$ relation of \cite{Maddox2015}. The combination of targets can be optimised based on redshift space, selection technique and detection probability. In addition there will be a significant increase in the number of lenses discovered in this redshift range in the period 2017-2023 from various multi-wavelength surveys, a {\sc Hi}-relevant summary of which is presented in \cite{Meyer2015}.

\begin{figure}
\centering
\includegraphics[width=0.8\textwidth]{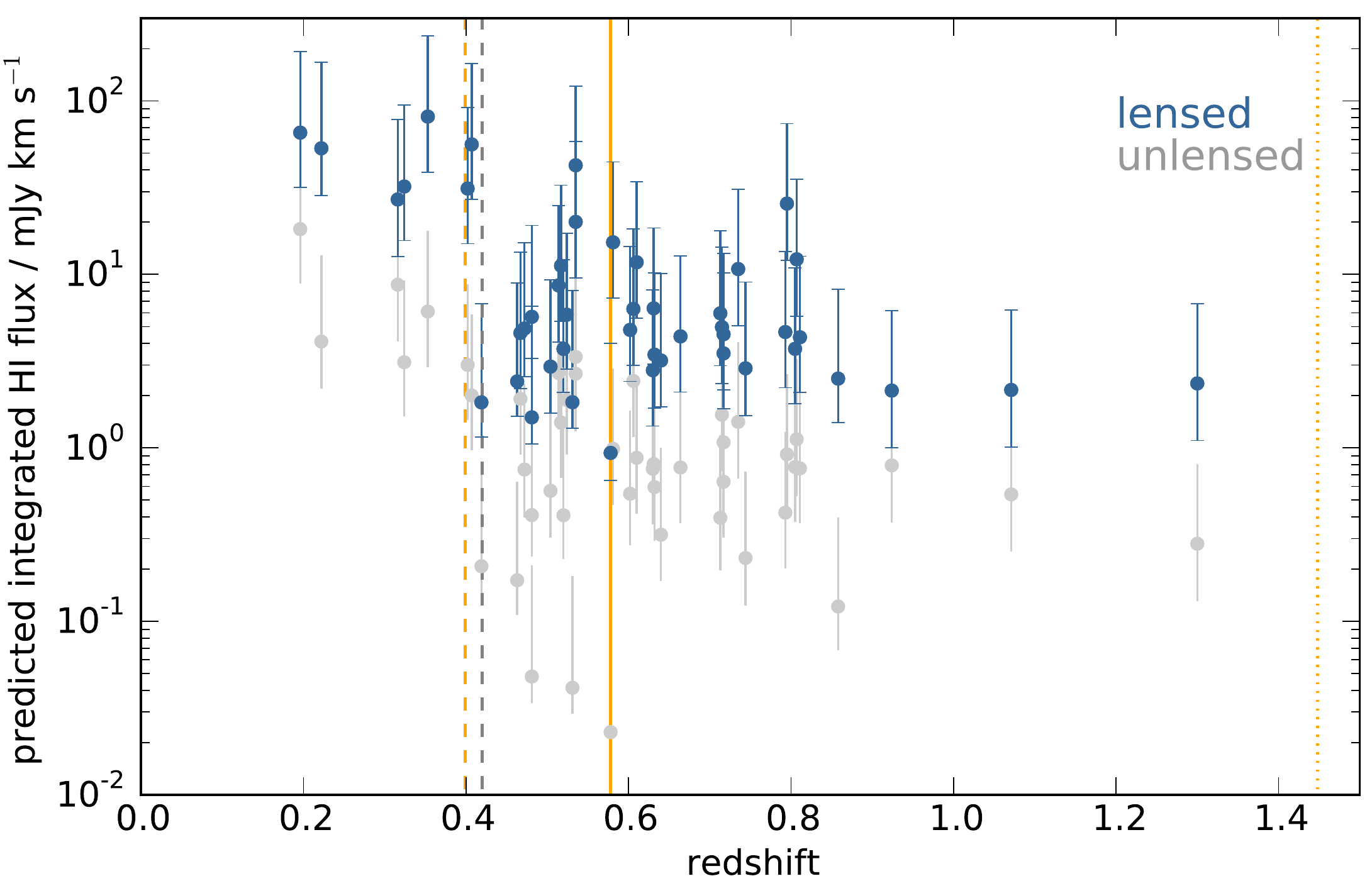}
\caption{Illustration of possible lensing targets from just one optical survey (SLACS) where all lens models were derived from HST imaging \cite{Newton2011}. Based on the optical magnifications, we plot both the predicted lensed (blue) and unlensed (grey) velocity-integrated \hi flux density. Reasonable assumptions about the velocity widths leads to a high probability of detections for many systems at the $\sigma \lesssim 50~\mu$Jy\,beam$^{-1}$ per 200 kHz channel sensitivity level (i.e. a few tens of hours of MeerKAT observing time). The dashed grey line indicates the maximum \hi redshift accessible by L-band receivers on current radio interferometers (e.g. VLA and GMRT). The orange dashed and solid lines indicate the MeerKAT UHF and L-band minimum and maximum redshift coverage respectively. The dotted orange line indicates the maximum \hi redshift accessible by the MeerKAT UHF band. }
\label{fig:knownlens}
\end{figure}

\subsection{Blind cluster surveys}\label{sec:cluster}

%need a cluster physics person: Anna Scaife? 

What is perhaps the most exciting and discovery-laden aspect of a potential \hi lensing programme, are deep observations of high mass clusters. The high projected mass surface density of a $z \gtrsim 0.1$ cluster dramatically increases the probability of lensed \hi detections. The Einstein radii of the highest mass clusters are all below $\theta_{\rm Ein} \lesssim 2$~arcmin, so well within the MeerKAT FoV. The strategy would be to target $z\sim0.3$ clusters with accurate mass models available (from optical weak/strong lensing and X-ray analyses), such as those in the Hubble Frontier Fields\footnote{See https://archive.stsci.edu/prepds/frontier/lensmodels for a detailed comparison of five different methods for six Hubble Frontier Field clusters.}. As discussed in Sec.~\ref{sec:design}, a deep integration in the UHF band would open up a lensed \hi discovery space across a co-moving cosmological volume of $\sim 15$~Gpc$^3$. An additional consideration would be to select clusters with minimal extended emission and hence more accurate sky models and resultant image fidelity, however, with 2016 baselines (for 64 antennas), modelling any complex source structure is a well-constrained problem and can be further circumvented by excluded the shortest baselines in the complex gain solution determination. 

The cluster-lens searches could be carried out in two tiers: several clusters observed in both L- and UHF band for 20-50 hours; and then the ultra-deep observation of 1-2 clusters in UHF for 100-250 hours. Based on simulations from \cite{Deane2015}, the latter will result in 1 guaranteed discovery beyond $z\sim1$, as well as several lower redshift and known lenses from optical searches. Although these data will be confusion-limited in total intensity, they will be deepest polarized view of galaxy clusters. With MeerKAT's optimised imaging fidelity, dynamic range and brightness temperature sensitivity, this could be a revolutionary project in itself and leave a lasting legacy into the SKA era. 

There are a wide range of polarisation analyses possible with such a rich dataset, including large-scale polarisation angle alignment, polarisation percentage as a function of cluster radius, polarisation stacking experiments, and \hi stacking of the cluster members themselves which, for the sake of brevity, we do not elaborate on here. %This is a ripe and unique role that MeerKAT could play. 

%With such promising potential, it would make sense to perform single, deep cluster observation in the UHF band, determine the results and decide from there. Shorter integration ($\sim$50 hours) on a wider range would make sense too. 

%basically, with all this stuff: give us a chance to demonstrate that we can detect this stuff.If yes, let's move forward. If no, you've only spent ~500 hours, a lot of which is still in commissioning time. 

%%%%%%
%highly complementary with targets survey of known lenses. 

%stacking possible

%Something that can be revolutionary in itself, as it will be a unique study of diffuse polarized emission clusters. All MeerKAT cluster surveys are just a few hours per pointing and at L-band to avoid confusion. Not confused in polarised intensity. 

\section{Complementarity with and enhancement of MeerKAT Large Survey Projects}

This contribution has focused on presenting the lensed \hi case with MeerKAT. For this reason, as well as brevity, we have not spent much time on the revolutionary impact that the MeerKAT LADUMA and MIGHTEE-HI surveys will have on our view and understanding of the \hi Universe. However, by carrying out an ambitious lensing programme, one can greatly enhance these surveys and almost consider \hi lensing as a third tier in the MIGHTEE-HI and LADUMA LSPs.

As discussed in Sec.~\ref{sec:targeted}, gravitational lensing enables the detection of systems of mass and at redshifts not directly accessible by LADUMA. This will not only add to the output of LADUMA, but also provide a very useful cross-check with stacking, provided a sufficient number of lensed \hi detections are made at high redshift. Lens modelling has made significant strides with a wide range of algorithms producing consistent macroscopic lens models for complex cluster morphologies (e.g. the Hubble Frontier Field lens models, as described earlier), lowering the associated systematic uncertainties. For large solid angle sources such as \hi, and for galaxy-galaxy lensing, this is further reduced.  Therefore, from a high-redshift \hi risk perspective, lensing circumvents most of the calibration risks that the deep LSPs will face and will return results on very short timescales.

High complementarity is expected with MeerKAT cluster surveys (e.g. Bernardi et al., Knowles et al., these proceedings). The dedicated MeerKAT cluster surveys are typically focused on mass or SZ-selected clusters over a wide range of redshifts. The objectives are primarily on the detection of radio halos and relics and are generally a few hours per source and therefore not deep enough to detect lensed {\sc H\,i}.  The emphasis in proposed cluster surveys is on driving up statistics up with large sample sizes, rather than observations of a depth capable of detecting lensed {\sc H\,i}. As such, deep integrations on high mass clusters as part of a dedicated \hi lensing programme will push MeerKAT's imaging fidelity and high dynamic range niche and open up cluster discovery space not possible with the current cluster survey prescription. Given the confusion noise limits, this primarily pertains to the polarized intensity. 

If a wide area ($\Omega \gtrsim 4,000$~deg$^2$) MeerKAT survey is performed, as considered elsewhere in these proceedings (Santos et al.), this should undoubtedly yield some extremely rare, low-redshift, high-magnification lensed \hi sources. While our 150~deg$^2$ simulation is not well-placed to make detection predictions for such a survey, and the band choice is to be determined, the sensitivity of MeerKAT is expected to open potential discovery space from a lensed \hi perspective. 

While these proceedings are focused on lensed H\,{\sc i}, this will naturally enable simultaneous detections of total and polarisation intensity of these sources, enabling detailed study of the star formation and ISM physics at high redshift.

%Big data idea: need to search for lensed HI behind every cluster and elliptical galaxy in all MeerKAT observations. Could do this at IDIA. Need to start MIGHTEE-LENS and and LADUMA-LENS. 

\section{Synergy with other major facilities and surveys}

We expect this lensed \hi programme to have strong synergies with other observatories and large-scale multi-wavelength imaging and spectroscopic surveys. This will be from both a target selection and detection follow up perspective. Over the course of the next $\sim$5 years, significantly more lenses will be discovered/confirmed in the wide area, optical/NIR imaging and spectroscopic programmes (see \cite{Meyer2015} for a summary), so the targeted lensed \hi component to this proposed programme would \emph{not} be target limited. Naturally, there will be significant synergies with SALT, with regard to obtaining spectroscopic or photometric redshifts of the foreground lens, as well as stellar mass constraints. Cross-correlation with these multi-wavelength data will also enable confirmation on marginal \hi detections. 

In the follow up context, there are a slew of multi-wavelength observatories that enable lensed \hi sources to each become individual cosmic laboratories in their own right. Firstly, followup with FAST will be highly synergistic with the lower mass, high magnification systems discovered in the cluster searches. FAST is highly sensitive but cannot act as a survey instrument for the depth and area required for blind, lensed \hi cluster searches. The compact MeerKAT core will also provide an accurate, minimally extrapolated continuum map for FAST in order to perform accurate continuum subtraction. 

There will be important follow up opportunities for both unresolved detections with the Large Millimetre Telescope in Mexico and potentially resolved galaxy kinematics with ALMA. This would not only enable mass modelling and dynamics, but also allow comparisons with the \hi spectra, as outlined in Sec.~\ref{sec:highz}. Finally, there is of course the potential JWST followup of the most spectacular systems.

\section{Pathway to and legacy value in the SKA era}

Following from MeerKAT, the SKA1-MID array will take lensed \hi studies forward in two primary ways:

\begin{enumerate}
\item The SKA1-MID factor $\sim3$ sensitivity increase and extended Band 1 redshift coverage to $z = 3$ will allow blind surveys to discover large samples of lensed \hi sources and not be limited to massive cluster searches. This will fully exploit the selection technique described in Sec.~\ref{sec:lensselect}. 
\item For the most massive galaxy clusters ($\theta_{\rm Ein} \gtrsim 20$~arcsec), it will become possible to spatially resolve intermediate redshift sources.
\end{enumerate}

A dedicated MeerKAT \hi lensing programme will provide immediate targets for SKA1-MID as well as guidance for optimised SKA cluster and blind lensed \hi surveys. In the full SKA era, lensed \hi may well be the most efficient lens selection tool, particularly when used in combination with the lensed continuum emission, \hi absorption/emission in the foreground lens, and OH lines in the same band (see McKean et al. 2015). 

%resolved studies of known lenses

%predictions on MW-like systems 

%guidance towards Band 1 commissioning? Just like SDP.81 was for ALMA. 

\section{Conclusion}

The revolutionary role MeerKAT has been designed to play in uncovering the \hi history of the Universe will be dramatically enhanced by exploiting the natural magnification afforded by strong gravitational lensing. In these proceedings, and in \cite{Deane2015}, we have made the case that:

\begin{enumerate}

\item MeerKAT will (emphatically) be the best facility to detect lensed \hi pre-SKA.
\item A dedicated MeerKAT \hi lensing programme will provide high-impact, rapid-turnaround early science and will make the highest redshift detections of \hi emission in galaxies pre-SKA. 
\item Direct lensed \hi detections at high-redshift will provide important cross-checks with stacking/statistical methods, provided a sufficient number of lensed-\hi detections are made.
\item The lensed \hi observing requirements pose significantly lower risk regarding calibration for high-z \hi inference, when compared to the LSPs.
\item As has been seen in the case of sub/mm-discovered lenses, MeerKAT-discovered \hi lenses are likely to have high legacy value well into SKA era.

\end{enumerate}

%Prospect are good. Great, niche opportunity for MeerKAT that can be easily tested in during science commissioning time. 

\section*{Acknowledgements}

\noindent We thank Natasha Maddox for allowing us to reproduce Figure~\ref{fig:maddox} here and for helpful discussions.

\end{document}